\documentclass[twocolumn,showpacs,prl]{revtex4}
\usepackage{amsmath}
\usepackage{graphicx}
\usepackage{dcolumn}
\usepackage{bm}

\begin{document}

\title{Delocalization in two dimensional disordered Bose
systems and depinning transition in the vortex state in
superconductors. }

\author{A.V. Lopatin and V.M. Vinokur}
\affiliation{Materials Science Division, Argonne National Laboratory, Argonne, Illinois
60439}
\date{\today}

\begin{abstract}
We investigate two-dimensional Bose system with the long range
interactions in the presence of disorder. Formation of the bound
states at strong impurity sites gives rise to an additional
depletion of the superfluid density $n_s$.  We demonstrate the
existence of the intermediate superfluid state where the
condensate and localized bosons present simultaneously.  We find
that interactions suppress localization and that with the increase
of the boson density the system experiences a sharp {\it
delocalization crossover} into a state where all bosons are
delocalized. We map our results onto the three dimensional system
of vortices in type II superconductors in the presence of columnar
defects; the intermediate superfluid state maps to an intermediate
vortex liquid where vortex liquid neighbors pinned vortices.  We
predict the {\it depinning transition} within the vortex liquid
and  {\it depinning induced} vortex lattice/Bose glass melting.
\end{abstract}

\maketitle

Effects of disorder on strongly correlated systems are central to
contemporary condensed matter physics. Disordered Bose systems are of
special interest \cite{Experiments,theory,Fisher}. Superfluid $^4$He and $^3$%
He remain one of the favorite and important experimental tools for studies
macroscopic quantum effects. Moreover, quantum mechanical mapping that
relates quantum Bose system to thermodynamics of superconducting vortices in
the space of the one dimension up, motivates research of Bose fluids in
random environment by the hope for further progress in vortex physics.

There has been a significant recent advance in understanding disordered Bose
systems started by \cite{Huang,Pitaevskii} who discussed a continuum model
of the dilute interacting Bose gas in a random potential. The proposed model
described the quasiparticle dissipation and depletion of superfluidity at
zero temperature. Later we developed a systematic diagrammatic perturbation
theory for the dilute Bose gas with weak disorder that enabled us to extend
the description of superfluid thermodynamics on finite temperatures \cite{LV}%
. Disorder corrections to the thermodynamic potential and the
disorder-induced shift of the condensation temperature resulting from
disorder scattering of quasiparticles were found. In particular, we obtained
that the superfluid density decreases monotonically with the temperature.

Yet there are many questions that remain open. First, more close examination
of models \cite{Huang,Pitaevskii,LV} shows that the described depletion of
superfluid density reflects only \textit{scattering} of zero energy
quasiparticles by random potential, while possible contribution from the
quasiparticles bound states was overlooked. Further, the fact that some
quasiparticles leave condensate and get localized suggests the existence of
the novel \textit{intermediate} state where superfluid and localized
components present simultaneously. This poses the next question: if this
intermediate state exists, what part of the phase diagram does it cover? In
other words the question is: would even arbitrarily weak disorder localize
the part of the condensate, or else this localization effect \textit{vanishes%
} if disorder is too weak or if the Boson interactions are sufficiently
strong?

In this Letter we examine the behavior of 2D interacting Bose-system subject
to strong disorder having in mind application of our results (via quantum
mechanical mapping) to the system of 3D superconducting vortices in type II
superconductors pinned by columnar defects. It is well established that at
low temperatures where the pristine vortex structure would form Abrikosov
lattice, columnar defects cause formation of strongly pinned Bose glass \cite%
{nelsvin,expBG}. At the melting line $B_m(T)$, Bose glass melts into a
vortex liquid. In the related quantum mechanical 2D system vortex liquid
corresponds to a superfluid Bosons. The depletion of the superfluid density
by disorder corresponds to enhancement of the average tilt modulus of the
related vortex system \cite{HNLDV,larkvin}. Moreover, one can expect that
this stiffening is accompanied by the change in transport properties: the
part of the vortices remain pinned and does not participate in transport.
This may be interpreted as the intermediate vortex phase where both vortex
liquid and pinned vortices coexist.

We focus on the most interesting case where the applied magnetic filed is
not too close to $H_{c1}$ such that distance between the vortices is less
than their interaction range, which coincides with the magnetic field
penetration length $\lambda.$ We show that upon increasing the temperature
(or, equivalently, magnetic field ), the system undergoes a sharp crossover
between the phase containing vortices pinned by the columnar defects and the
phase where all vortices are delocalized and find the expression for the
crossover line:
\begin{equation}  \label{result}
\rho={\frac{{c\,\kappa^4 }}{{\ \l _T^2\ln(\lambda/L)}}}\, \exp \left[ - {%
\frac{{\ \sqrt{8}\, \lambda^2 }}{{\epsilon \,r_r l_T }}} \right],
\end{equation}
where $\rho$ is the density of vortices, $\kappa=\lambda/\xi \gg 1 $ is the
ratio of the magnetic penetration and coherence lengths, $r_r$ is the
characteristic size of the columnar defect, $\epsilon$ is the anisotropy
parameter, the length $l_T$ is determined by the temperature $T$ as $%
l_T=(\Phi_0/4\pi)^2/T, $ and $c$ is a numerical coefficient of the
order one. The localization length $L$ that enters the logarithmic
factor is defined below in Eq. (\ref{loc_length}). The
corresponding phase diagram is shown on Fig. 1 where the
delocalization line is drawn along with the vortex lattice melting
line. In what follows we formulate our model in terms of vortex
configuration, establish the existence of the delocalization
crossover, and derive equation (\ref{result}) for the crossover
line.

\textit{Model.} A free energy of the vortex configuration can be written as
a functional of the vortex displacements $\mathbf{r}_i (z)$ as
\begin{equation}
F=\int^L_0 dz \left[\sum_i {\frac{\epsilon_1 }{2}} \left({\frac{{\partial
\mathbf{r}_i }}{{\partial z}}} \right)^2 +U(\mathbf{r}_i)+{\frac{1}{2}}
\sum_{i,j} v(\mathbf{r}_{ij}) \right]
\end{equation}
where the interaction between the vortices is
\begin{equation}
v(r)=\epsilon_0\, K_0(r/\lambda),
\end{equation}
with $\epsilon_0=(\Phi_0/4\pi\lambda)^2$ and $K_0$ being the Bessel
function. The vortex tension $\epsilon_1$ is related to the anisotropy
parameter $\epsilon$ and $\epsilon_0$ as $\epsilon_1=\epsilon^2 \epsilon_0$
and the disorder potential $U(r)$ can be written as a sum of the potential
wells representing the columnar defects
\begin{equation}
U(r)=\sum_i \, u(r-r_i),
\end{equation}
where $r_i$ is the coordinate of the $i-th$ defect.

Statistical mechanics of vortices can be mapped onto the quantum mechanical
problem of interacting two-dimensional bosons described by the Hamiltonian
\begin{eqnarray}  \label{Hamiltonian}
\hat H &=& \int d^2 r\, \hat \psi^\dagger (r)\,[\hat p^2/2m-\mu+U(r)] \,\hat
\psi(r)  \notag \\
&&+ {\frac{1}{2}}\, \int d^2 r_1 d^2 r_2 \; \hat n(r_1) \, v(r_1-r_2)\, \hat
n(r_2),
\end{eqnarray}
where $\hat \psi^\dagger ,\hat \psi$ are bose creation and annihilation
operators, $\mu$ is the chemical potential and $\hat n(r)=\hat
\psi^\dagger(r)\, \hat \psi(r) $ is the density operator of the bose gas.
Results obtained in the Bose gas representation can be translated onto
vortex language by the substitute $\hbar\to T,\, m\to \epsilon_1.$ The
energy of the Bose gas corresponds to the free energy per unit length of the
vortex lattice. Keeping in mind the application to vortices we will make
several assumptions about the form of a single potential well: We will
consider temperatures $T\approx T_c, $ in this regime pinning potential is
given by
\begin{equation}  \label{potential_well}
u(r)= {\frac{{\epsilon_0}}{2}} \,{\frac{{r_r^2 }}{{r^2+2\xi^2 } }},
\end{equation}
where $\xi$ is the coherence length and $r_r$ is the scale that
characterizes the size of the columnar defect. The strength of the pinning
potential can be characterized by the dimensionless coefficient
\begin{equation}
\beta=r_r^2 \epsilon_1\epsilon_0/2T^2.
\end{equation}
We will assume that pinning is weak such that $\beta\ll 1.$ In this limit
the solution of the Schrodinger equation
\begin{equation}
[-\nabla_r/2\epsilon_1 +u(r)\,]\, \psi(r)=E_1\, \psi(r),
\end{equation}
results in the pinning energy
\begin{equation}  \label{eps_1}
E_1\sim {\frac{{T^2 }}{{\ \epsilon_1 \xi^2 }}}\, e^{-1/\sqrt{\beta} }.
\end{equation}
The localization length of the quantum sate is related to the bound state
energy as $L^2 \sim \hbar^2/|E_1|m,$ translating this relation on vortex
language we define the localization length of the vortex as on the columnar
defect as
\begin{equation}  \label{loc_length}
L = \xi \, e^{1/2\sqrt{\beta} }.
\end{equation}
Further, we assume that the interaction between the bosons (vortices) is
strong enough such that the bound state of the well can be only single
occupied. This condition is satisfied when the energy of the double occupied
state $E_2$ is much larger than the the absolute value of the energy of the
single occupied state $E_1.$ The interaction energy of the double occupied
state is
\begin{equation}
E_2={\frac{1}{2}}\, \int d^2 r_1 d^2 r_2 \; \phi_1^*(r_1)
\phi_1(r_1) \, v(r_1-r_2)\, \phi_1^* (r_2) \phi_1(r_2),
\end{equation}
where $\phi_1$ is the wave function of the bound sate. In the case when the
localization length $L$ is smaller than the interaction range $\lambda$ the
energy $E_2$ can be estimated as
\begin{equation}
E_2\sim \epsilon_0\, \ln(\lambda/L),
\end{equation}
where we have used the asymptotic of the Bessel function $K_0(x) \approx \ln
(1/ x).$ The condition of no double occupation can be now written as
\begin{equation}  \label{no_double_oc}
\epsilon_0 \epsilon_1 \, \ln (\lambda / L ) \gg (T/L)^2 .
\end{equation}
In case of low concentration of defects we can consider the contributions of
different potential wells separately, thus we end up with the problem of the
interacting Bose gas and single potential well ( or a single columnar defect
placed in the vortex liquid). Leaving only one potential well we can easily
diagonolize the non-interacting Hamiltonian and formulate the problem in
terms of eigenfunctions of the noninteracting Hamiltonian
\begin{equation}
\hat \psi = \sum \hat b_k\, \phi_k,
\end{equation}
where $\phi_k$ are the eigenfunctions of the non-interacting Hamiltonian
\begin{equation}
[\, \hat p^2/2m+u(r)]\, \phi_k(r)= E_k \, \phi_k(r),
\end{equation}
and $E_k$ is the energy of the $k-th$ state. The bound state corresponds to
the index $k=1.$ Although this state has the lowest energy, the
Bose-condensation to this state cannot occur due to our assumption of
no-double occupation. Thus, the condensation takes place to the extended
state with the lowest energy (which we label with $k=0).$ The normalization
conditions for eigenfunctions $\phi_k$ are as follows:
\begin{equation}
\int d^2 r \, \phi^*_k\phi_k=\left \{
\begin{array}{c}
V, \;\;\;\; E_k >=0 \\
1, \;\;\;\;\;k=1%
\end{array}
\right.
\end{equation}
with $V$ being the volume. In the absence of the current the basis can be
chosen real, thus we will assume further on that $\phi^*=\phi.$ The $\psi$%
-operators that enter the Hamiltonian (\ref{Hamiltonian}) can be presented
as a sum of three terms
\begin{equation}  \label{psi}
\hat \psi=\phi_0\, \hat b_0 +\phi_1 \hat b_1+ \sum_{E_k>0}\,
\phi_k\, \hat b_k,
\end{equation}
representing the condensate, the bound state, and all other states having
energy larger than the condensate energy. The effective temperature of the
Bose gas is related to the inverse physical length of the superconducting
sample in the $z$ direction. We will assume that the size of the sample is
microscopically large, this corresponds to zero temperature limit for
bosons. We will consider the most interesting case where the applied
magnetic filed is not to close to $H_{c1},$ such that the average distance
between the vortices is smaller than the screening length $\lambda.$ In this
regime the mean filed Bogolubov approximation, that we will use, is
justified for the case of sufficiently high density of vortices. \cite{FGIL}
Thus, the third term in Eq.(\ref{psi}) representing the particles that do
not belong to the condensate is small and can be omitted in the leading
order.

Now we turn to the analysis of the population of the bound state in the
presence of the condensate. Inserting the representation Eq.(\ref{psi}) into
the Hamiltonian (\ref{Hamiltonian}) and keeping only two first terms in the
representation (\ref{psi}) we get the effective Hamiltonian
\begin{equation}  \label{H_eff}
\hat H_{eff}=b_1^\dagger\, (E_1+\alpha-\mu)\, \hat b_1 + \gamma \,
(\hat b_1^\dagger+\hat b_1),
\end{equation}
where the terms $\hat b_1^\dagger \hat b_1^\dagger$ were not written due to
no double occupancy condition. The coefficients $\alpha$ and $\gamma$ in the
Hamiltonian (\ref{H_eff}) are
\begin{eqnarray}  \label{alpha}
\alpha=b_0^2 &\int& d^2 r_1 d^2 r_2 [\phi_1^2(r_1)\, v(r_1-r_2) \phi_0^2(r_2)
\notag \\
&+& \phi_1(r_1) \phi_0(r_1)\, v(r_1-r_2) \phi_1(r_2) \phi_0(r_2)],
\label{alpha} \\
\gamma = b_0^3 &\int& d^2 r_1 d^2 r_2 \, \phi_0^3(r_1)\,
v(r_1-r_2) \, \phi(r_2),  \label{gamma}
\end{eqnarray}
where the average value $b_0$ of the operator $\hat b_0$ is related to the
condensate density $\rho_0$ as $b_0^2=\rho_0.$ The model (\ref{H_eff}) can
be easily solved: Presenting the wave functions in the form
\begin{equation}
\psi=a_0 |0> +a_1 |1>,
\end{equation}
with $a_0$ and $a_1$ being the amplitudes of the zero and single occupied
states we find two eigenstates corresponding to the energy levels
\begin{equation}
E_\pm={\frac{{E \pm \sqrt{E^2+4\gamma^2}}}{{2 }}}
\end{equation}
where $E=E_1+\alpha-\mu.$ We see that $E_+>0$ and $E_- <0$ for any sign of
the energy $E,$ thus the state with energy $E_-$ is always occupied while $%
E_+$ is always empty. The occupation of the center is determined by the
state with the lowest energy ($E_-$) and is given by
\begin{equation}  \label{ocupation}
n=a_1^2={\frac{{2
}}{{4+(E/\gamma)^2+(E/\gamma)\sqrt{(E/\gamma)^2+4} }}}.
\end{equation}
One can easily see that $n \to 1$ when $E/\gamma\ll -1$ and $n\to 0$ when $%
E/\gamma \gg 1.$ Thus the quantity $E/\gamma$ controls the
occupation such that at $E=0$ localization-delocalization
crossover takes place. Making use of Eq.(\ref{alpha}) and
relationship between the chemical potential, condensate density
and interaction $\mu=b_0^2 v_0$ where $v_0=\int d^2 r \, v(r),$
for the parameter $E$ we find $E=E_1+\delta E$ with
\begin{equation}  \label{deltaE}
\delta E=b_0^2 \, \int d^2 r_1\, d^2 r_2\, \phi_1(r_1)\, v(r_1-r_2)\,
\phi_1(r_2) \approx \rho L^2\, \epsilon_0\, \ln (\lambda/L )
\end{equation}
Now, using the relationship between the bound state energy and
localization length, from the condition $E=0,$ we get the vortex
density at the localization-delocalization crossover
\begin{equation}  \label{rho}
\rho= {\frac{{\ c\, T^2}}{{\epsilon_0\, \epsilon_1\, L^4 \,
\ln(\lambda / L)} }},
\end{equation}
where $c\sim 1$ is a numerical constant. Plugging in the localization length
from Eq. (\ref{loc_length}) 
we arrive at the expression (\ref{result}) for the delocalization crossover
line which can be rewritten in terms of the \textit{depinning field} for
vortices as
\begin{equation}  \label{dep_line}
B_{dep}\simeq \Phi_0 \frac{2\pi c  T^2}{\epsilon_0\epsilon_1\xi^4\ln(\lambda/L)%
} \exp\bigg(-\frac{T}{T_0}\bigg),
\end{equation}
where $T_0$ is the effective temperature dependent depinning
energy $T_0=r_r\sqrt{\epsilon_1\epsilon_0/8}.$ One can easily
check that the assumed condition of no double occupancy
(\ref{no_double_oc}) always satisfies near the crossover as long
as $\rho L^2 < 1.$

\begin{figure}[ht]
\includegraphics[width=3.2in]{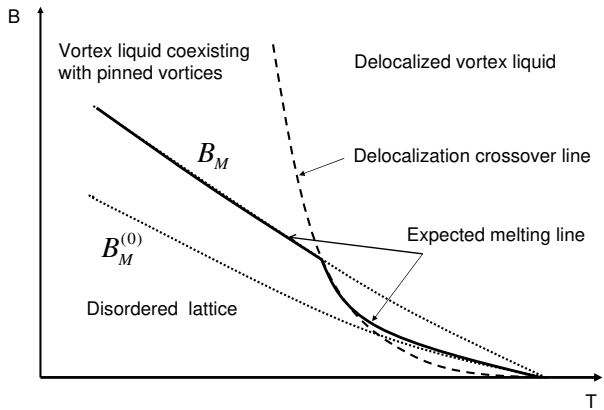}
 \caption{ Schematic phase diagram for the vortex system with columnar defects.
 The melting line under assumption of pinning is denoted by $B_M,$ while the melting line of the
 pristine lattice is denoted by $B_M^{(0)}.$  }
 \end{figure}

As we see from Eq.(\ref{ocupation}), the occupation of the center
changes continuously as long as ``hybridization parameter"
$\gamma$ is larger than zero. Therefore the phase boundary between
the localized and delocalized phases should be viewed as a
crossover rather than the true transition line. Note, however,
that in case of finite concentration of defects the delocalization
transition may result in significant change of the condensate
density, that may induce the true superfluid-bose glass transition
studied in \cite{larkvin}.


While the ultimate type of the crossover remains an open question,
experimentally it can appear very sharp and undistinguishable from
the true first order transition. Indeed, the sharpness of the
crossover is controlled by the ratio of $\gamma$ and $\delta E.$
Formula (\ref{gamma}) determining $\gamma,$ should be corrected to
include the screening of the potential by the surrounding Bose
gas. In the dense limit the screening length is given
by $l_{sc}=\lambda (1/2\pi \rho \epsilon^2 \,l_T^2 )^{1/4}.$ The parameter $%
\gamma$ is estimated then as
\begin{equation}
\gamma\approx {\frac{{\rho L \lambda^2 \epsilon_0}}{{\ l_T \epsilon \, \sqrt{%
2\pi}}}},
\end{equation}
and the ratio $\gamma/\delta E$ becomes:
\begin{equation}
\gamma/\delta E \sim \lambda^2/(l_T L\, \epsilon\,\ln(\lambda/L))
\end{equation}
and is always small due to the no-double occupancy condition (\ref%
{no_double_oc}). We thus conclude that the crossover between the localized
and non-localized phases is sharp.

We are now in a position to construct a phase diagram for the disordered
Bose system. We use the corresponding `superconductor vortex language'; the
schematic vortex phase diagram for the system with columnar defects is
presented in Fig 1. Shown are the melting line of the pristine sample $B_{%
\scriptscriptstyle M}^{(0)}$, and melting line $B_{%
\scriptscriptstyle M}$ in the sample with the columnar defects
shifted, as it should \cite{nelsvin, expBG}, upwards as compared to $B_{%
\scriptscriptstyle M}^{(0)}$. Note that because of the exponential
dependence of temperature, $B_{dep}$ drops much faster than $B_{%
\scriptscriptstyle M}$ as temperature approaches $T_c$.
Thus the depinning line that lies in the vortex liquid domain at high
sufficiently fields hits and then crosses the melting line $B_{%
\scriptscriptstyle M}$, traversing then to $T_c$ and crossing pristine
melting line at very low fields.

In the liquid phase the depinning line marks the crossover
separating the intermediate state with the columnar defects
occupied by vortices (i.e. vortex liquid coexists with the pinned
vortices), and the phase where all vortices are delocalized from
the columnar defects. In the strip
confined between the $B_{\scriptscriptstyle M}^{(0)}$ and $B_{%
\scriptscriptstyle M}$ lines, $B_{dep}$ marks melting of the
vortex lattice. Indeed, at $B_{\scriptscriptstyle
M}^{(0)}<B<B_{\scriptscriptstyle M}$ lattice is stabilized by
vortices pinned by columnar defects. As soon as depinning at
$B_{dep}$ occurs, columnar defects become inessential, and lattice
looses its stability since in the absence of defects vortex liquid
is a stable thermodynamic state above $B_{\scriptscriptstyle
M}^{(0)}$. We therefore discover a novel type of the vortex lattice melting,
the \textit{depinning induced melting}, that occurs in the interval $B_{%
\scriptscriptstyle M}^{(0)}<B<B_{\scriptscriptstyle M}$.

This new kind of vortex lattice melting may well explain the
origin of the low-field kink in the melting line that clearly
indicates a switch between the different melting mechanisms. This
kink has been observed in almost all experiments on the Bose glass
melting \cite{expBG}, and was addressed specifically in the recent
study of the melting of ``porous'' vortex matter \cite{eli}.
Although these results obtained for pronounced vortex lines may be
applied only with caution and many reservations to BSCCO, it is
interesting to note that at temperatures around 80K the
characteristic energy $T_0$ for BSCCO parameters can be estimated
as $T_0\simeq 5K$, close enough to that observed in \cite{eli}.

In conclusion we have shown that in a disordered Bose system with
the long range interactions an intermediate phase where both
superfluid and localized bosons can exist.  We demonstrated that a
sharp delocalization crossover occurs with the increase of bosons
density.
 We apply our results to description of a vortex system in type II
superconductors in the presence of columnar defects and show that
this delocalization crossover describes the depinning line
separating two liquid phases, the intermediate phase containing
pinned vortices and 'fully molten' liquid.  We predict a new kind
of the first order vortex lattice melting, depinning induced
melting, at low fields.

It is a pleasure to thank S. Banerjee, A. Koshelev, D. Maslov, and
E. Zeldov for illuminating discussions.  This work was supported
by the U.S. Department of Energy Office of Science through
contract No. W-31-109-ENG-38.

\end{document}